\documentclass[12pt]{article}
\usepackage{a4wide}
\usepackage[centertags]{amsmath}
\usepackage{amssymb}
\usepackage[sort&compress,numbers]{natbib}
\usepackage{ifpdf}
\usepackage{setspace}
\usepackage[dvips]{graphicx} 
\usepackage{amsfonts}
\usepackage{subfigure}
\usepackage{threeparttable}
\ifpdf
\usepackage[bookmarks]{hyperref}
\else
\usepackage[hypertex]{hyperref}
\usepackage{epsfig}
\usepackage{latexsym,amsmath,amssymb,amstext}
\fi
\newcommand{\half}{\frac{1}{2}}

\usepackage[dvips]{color}
\usepackage{feynmp}
\def\Tr{\mathrm{Tr}}

\def\half{{1\over2}}
\def\nn{\nonumber\\}

\def\[{\left[}
\def\]{\right]}
\def\({\left(}
\def\){\right)}

\newcommand{\wt}{\widetilde}

\def\={\stackrel{\bullet}{=}}
\def\nn{\notag\\}

\def\Tr{\mathrm{Tr}}

\def\half{{1\over2}}

\def\[{\left[}
\def\]{\right]}
\def\({\left(}
\def\){\right)}

\def\cO{{\cal O}}

\def\bS{{\mathbf S}}

\def\bR{{\mathbf R}}
\def\bZ{{\mathbf Z}}

\def \be {\begin{equation}}
\def \ee {\end{equation}}
\def \bea {\begin{eqnarray}}
\def \eea {\end{eqnarray}}
\def \beal#1 {\begin{align}#1\end{align}}
\def \nn {\notag\\}

\def\aver#1{\left\langle #1 \right\rangle}

\def\beal#1{\begin{align}#1\end{align}}

\begin{document}

\setlength{\topmargin}{-.1in} 
\setlength{\textheight}{9.0in}
\makeatletter
\renewcommand{\theequation}{%
\thesection.\arabic{equation}}
\@addtoreset{equation}{section}
\makeatother

\begin{titlepage}
\vspace{-3cm}
\title{
\begin{flushright}
\normalsize{ TIFR/TH/12-34\\
Oct 2012}
\end{flushright}
       \vspace{1.5cm}
Chern-Simons-Fermion Vector Model \\
with Chemical Potential
       \vspace{1.5cm}
}
\author{
Shuichi Yokoyama
\\[25pt] 
\!\!\!\!\!\!\!\!{\it \normalsize Department of Theoretical Physics, Tata Institute of Fundamental Research,}\\
\!\!\!\!\!\!\!\!{\it \normalsize Homi Bhabha Road, Mumbai 400005, India}\\
\\[10pt]
\!\!\!\!\!\!\!\!{\small \tt E-mail: yokoyama(at)theory.tifr.res.in}
}

\date{}

\maketitle

\thispagestyle{empty}

\vspace{.2cm}

\begin{abstract}
\vspace{0.3cm}
\normalsize

We study SU($N$) level $k$ Chern-Simons theories coupling to a fundamental fermion with chemical potential for U($1$) flavor symmetry.   
We solve this system exactly in the 't Hooft limit, $N,k\to\infty$ with $\lambda = N/k$ fixed. 
The solution exists up to $|\lambda|=1$ for a fixed value of chemical potential. 
We study the system beyond $|\lambda|=1$ by considering U($N$) gauge group and taking the diagonal U($1$) holonomy into account. 
The result suggests the system becomes unstable and a phase transition happens at $|\lambda|=1$. 

\end{abstract}

\end{titlepage}


\section{Introduction}
\label{intro}

Vector (or sigma) models play important roles to understand not only phenomenology but also theoretical high energy physics. In the context of AdS/CFT duality they play a crucial role for the duality proposed in \cite{Klebanov:2002ja}. 
The duality claims that the singlet sector of a three dimensional O($N$) (or U($N$)) vector model with conformal symmetry and a higher spin gravity theory on AdS$_4$ \cite{Vasiliev:1990en,Vasiliev:2003ev,Bekaert:2005vh} in the large $N$ limit. This conjecture has been tested by comparing spectra and correlation functions of both sides \cite{Sezgin:2002rt,Sezgin:2003pt,Giombi:2010vg,Giombi:2011ya}.

Motivated by this conjecture three dimensional vector models have been studied in presence of CS interaction by gauging global symmetry \cite{Giombi:2011kc,Aharony:2011jz}. 
It was studied that conformal symmetry and higher spin symmetry are still preserved up to $1/N$ correction and 
the effect of CS interaction is reflected as a suitable change of boundary condition for a bulk scalar field. 
This class of conformal field theories turned out to have strongly constrained correlation functions \cite{Giombi:2011rz,Maldacena:2011jn,Maldacena:2012sf,Aharony:2012nh}.
The duality can be also studied by changing a base manifold \cite{Shenker:2011zf,Hellerman:2012} or in accordance with supersymmetry \cite{Sezgin:2002rt,Chang:2012kt,Sezgin:2012ag,Jain:2012qi}.  
A recent review is \cite{Giombi:2012he}.

In this paper we consider a little generalization of \cite{Giombi:2011kc} by adding chemical potential for U($1$) flavor symmetry into Chern-Simons-fermion vector model. 
One of the achievements in \cite{Giombi:2011kc} is solving the system by taking the light-cone gauge and the 't Hooft limit. The free energy is determined exactly as a function of $\lambda$. 
Interestingly the solution exists up to $|\lambda|=1$ and the free energy vanishes at that point. 
The motivation of this paper is to compute other various thermodynamic quantities such as the grand canonical potential and the thermodynamic descendants thereof exactly as a function of $\lambda$ and to address what happens to the system beyond $|\lambda|=1$.

The rest of the paper is organized in the following manner. 
In Section \ref{zerotemp} we present the exact solution of Chern-Simons-fermion vector model with chemical potential in the 't Hooft limit. 
In Section \ref{finitemp} we calculate various thermodynamic quantities exactly as a function of $\lambda$. 
In Section \ref{instability} we study this system beyond $|\lambda|=1$ by considering the gauge group  U($N$) and taking account of the diagonal U($1$) holonomy. The final section is devoted to summary and discussion.

\paragraph{Note added} 
The results of this paper should be corrected so that 
the nontrivial holonomy distribution found in \cite{Aharony:2012ns} is taken into account.

\section{Chern-Simons-fermion vector theory with chemical potential}
\label{zerotemp}

In this section we study SU($N$) Chern-Simons gauge theory at level $k$ with a fermion which is in the fundamental representation of the gauge group in presence of chemical potential $\mu$ for U($1$) flavor symmetry. 
The action of this system is 
\beal{
S  &= \int d^3 x  \biggl[i \varepsilon^{\mu\nu\rho} {k \over 4 \pi}
\Tr( A_\mu\partial_\nu A_\rho -{2 i\over3}  A_\mu A_\nu A_\rho)
 + \bar\psi \gamma^\mu D_\mu \psi  - \mu \bar\psi \gamma^3 \psi \biggl]
\label{originalaction}
}
We use the same convention in \cite{Giombi:2011kc,Jain:2012qi}.

This sytem can be exactly solved in the 't Hooft limit \cite{Giombi:2011kc}
\be
N, k \to \infty \quad \mbox{with}\quad \lambda:= {N / k}\quad \mbox{fixed}. 
\ee
The method is the following. Firstly one fixes the gauge degrees of freedom by the light-cone gauge in Euclidean space $A_- := {1\over \sqrt2}(A_1+i A_2)=0$ and integrate out the other components of the gauge field. 
This generates four fermi interaction. 
Secondly one introduces a bilocal singlet field $\Sigma$ so that the four fermi interaction term cancels, which is known as Hubbard-Stratonovich transformation. $\Sigma$ turns out to coincide with the self energy of the fermionic field on shell.
The resulting action is quadratic with respect to the fermion and 
thus we can integrate it out by gaussian integration. The remaining action is written in terms of the singlet field and $N$ can be factored out from the action. Finally one can obtain the exact effective action in the leading of $1/N$ expansion by the steepest descent method under the assumption of translational invariance.

From the action \eqref{originalaction} 
we can easily see that the result in our case can be obtained from that in \cite{Giombi:2011kc} by exchanging the momentum in the fermion propagator in such a way that 
\be
p_\mu \to \wt p_\mu := p_\mu+i\mu \delta_{\mu,3}.
\ee
After this repalacement we find the exact effective action 
\be
S_{eff} 
=NV \int \frac{d^3 q}{(2 \pi)^3} {\rm tr}
\left[- \log\left[ i \gamma^\mu \wt q_\mu +\Sigma(q) \right] 
+\frac{1}{2} \Sigma(q)  \left( \frac{1}{i \gamma^\mu \wt q_\mu + \Sigma(q)} \right)  \right]
\label{seff}
\ee
where the trace is taken over a two-by-two matrix with spinor indices and $V$ is the volume of the Euclidean space.
As shown in \cite{Giombi:2011kc}  
this effective action correctly reproduces the connected vacuum diagram consisting of planar graphs perturbatively with respect to $\lambda$. 
The exact propagator of fermionic field is
\be
\aver{\bar\psi(-p) \psi(q)} = {-1 \over i\gamma^\mu \wt p_\mu + \Sigma} (2 \pi)^3 \delta^3(-p+q).
\label{propagator}
\ee
Spinor indices are suppressed here and hereafter. 
We also find the saddle point equation 
\beal{
\Sigma(p) =& \int {d^3 q \over (2 \pi)^3} \biggl(\gamma^+ {1\over i \gamma^\mu \wt q_\mu   + \Sigma (q) }\gamma^3 - \gamma^3 {1\over i \gamma^\mu \wt q_\mu   + \Sigma (q) }\gamma^+ \biggl)  {2\pi \lambda \over i(p-q)_- }. 
\label{sd}
}
where $p_{\pm}={1\over \sqrt2}(p_1\mp ip_2),\, \gamma^{\pm} = {1\over \sqrt2}(\gamma_1\pm i\gamma_2)$. 
We can check that  
this equation precisely agrees with Schwinger-Dyson equation in the 't Hooft limit and also coincides with the equation of 1PI self-energy diagram, which is now given only by the rainbow diagram  \cite{Giombi:2011kc}.

Let us solve the saddle point equation \eqref{sd}. 
For this end we decompose $\Sigma$ into  
\bea
\Sigma =  \gamma^\mu \Sigma_\mu + I \Sigma_I 
\label{sigmadec}
\eea
where $I$ is a two by two unit matrix. 
Plugging this into \eqref{sd} we find
\bea
 ( \gamma^\mu \Sigma_\mu + I_2 \Sigma_I )(p) ={4\pi \lambda \over i } \int {d^3 q \over (2 \pi)^3} \biggl( { i (\wt q -i \Sigma)_- + \gamma^+  \Sigma_1 \over   (\wt q_\mu -i  \Sigma_\mu)^2 + \Sigma_1^2}  \biggl)  {1 \over (p-q)_- }
\eea
By comparing both sides we obtain 
\bea
&&\Sigma_-=\Sigma_3=0, \\
&&\Sigma_I(p) = {4\pi \lambda \over i } \int {d^3 q \over (2 \pi)^3} \biggl( { i q_-  \over   (\wt q _\mu - i  \Sigma_\mu(q))^2 + \Sigma_1(q)^2}  \biggl) {1 \over (p-q)_- } \\
&&\Sigma_+(p)=  {4\pi \lambda \over i } \int {d^3 q \over (2 \pi)^3} \biggl( {  \Sigma_I(q) \over   (\wt q_\mu -i \Sigma_\mu(q))^2 + \Sigma_1(q)^2}  \biggl)  {1 \over (p-q)_- }
\eea
From the dimensional analysis and rotational covariance in two plane we can set 
\bea
\Sigma_I(p) =  f(p') p_s, \quad \Sigma_+ (p) =i g(p') p_+
\label{sigmafg}
\eea
where $f$ and $g$ are unknown functions of $p' = {p_s \over \mu}$. 
Plugging these into the above equations we obtain 
\bea
&&f(p') p_s = 4\pi \lambda \int {d^3 q \over (2 \pi)^3} \biggl( {  q_-  \over   (1+g(q')) q_s^2 + \wt q _3^2 + (f(q') q_s ) ^2}  \biggl) {1 \over (p-q)_- }
\label{f} \\
&& g(p') p_+ = - 4\pi \lambda \int {d^3 q \over (2 \pi)^3} \biggl( {   f(q') q_s \over  (1+g(q')) q_s^2  + \wt q_3^2 + (f(q') q_s)^2}  \biggl)  {1 \over (p-q)_- }
\label{g}
\eea
By using 
$
{\partial \over \partial p_+} {1 \over (p-q)_-} =2\pi\delta^2 (p-q), 
$
we can show 
$
{\partial \over \partial p_+} \((f(p') p_s)^2 + g(p') p_s^2\) = 0, 
$
which leads to
\be
f(p')^2 + g(p') = {c_0 \over p'^2}, 
\label{fgc}
\ee
where $c_0$ is a dimensionless constant which will depend on $\lambda$. $g$ can be determined by \eqref{fgc} so we focus on solving the equation of $f$, \eqref{f}. 
Substituting \eqref{fgc} into \eqref{f} we obtain 
\bea
&&f(p') p_s = 4\pi \lambda \int {d^3 q \over (2 \pi)^3} \biggl( {  q_-  \over q_s^2 + \wt q _3^2 + c_0\mu ^2}  \biggl) {1 \over (p-q)_- }
\eea
Integrating the angular part of $p_-$ we get 
\bea
f(p') p_s = 4\pi \lambda\int_{p_s}^\infty {q_s d q_s \over 2 \pi}  \int {dq_3 \over 2 \pi}{  -1 \over q_s^2 + \wt q _3^2 + c_0\mu ^2} 
\eea
We regularize UV divergence by dimensional regularization, as done in \cite{Giombi:2011kc,Jain:2012qi}, which leads to%
\footnote{We note a usuful formula to derive this equation 
\bea
\int_\bR {d q_3 \over (q_3+i\mu)^2+a^2} = { \pi \over |a|} \theta(|a|- |\mu|). 
\eea
}
\beal{
f(p') 
&= { \lambda \over p'} \biggl( \sqrt{|p'^2+c_0| }\theta(\sqrt{|p'^2 +c_0| } - 1)  +  \theta(1-\sqrt{|p'^2 +c_0| } ) \biggl),
\label{f2}
}
where $\theta(x)$ is a step function taking unit value when $x>0$ otherwise zero.

$c_0$ can be determined as follows. 
First let us plug \eqref{f2} into \eqref{fgc} and expand $g$ around $p'\sim0$:  
\be
g(p') \sim {c_0 \over p'^2} -  { \lambda^2 \(\sqrt{|c_0| }\theta(\sqrt{|c_0| } - 1)  + \theta(1-\sqrt{|c_0| } ) \)^2 \over p'^2}.  
\ee
From \eqref{g} and \eqref{f2}, on the other hand, we can show $g(p')\sim0$ around $p'\sim0$ in the same way as in \cite{Giombi:2011kc,Jain:2012qi}. Therefore $c_0$ has to satisfy
\be
c_0 - \lambda^2 \(\sqrt{|c_0| }\theta(\sqrt{|c_0| } - 1)  +  \theta(1-\sqrt{|c_0| }) \)^2 =0
\ee
A unique solution exists only when $|\lambda|<1$ as $c_0 =\lambda^2$ and no solution when $|\lambda|>1$.

As a result we obtain the solution of the saddle point equation 
\beal{
f(p') &= { \lambda \over p'} \biggl( \sqrt{ p'^2+\lambda^2 }\theta(\sqrt{ p'^2 +\lambda^2 } - 1)  +  \theta(1-\sqrt{p'^2 +\lambda^2 } ) \biggl),
\label{solf}\\
g(p') &= { \lambda^2 \over p'^2} (1- p'^2-\lambda^2 ) \theta(\sqrt{ p'^2 +\lambda^2 } - 1),
\label{solg}
}
where $|\lambda|<1$. $\Sigma$ finally becomes  
\be
\Sigma(p) = f(p') p_s I + i g(p') \gamma^+ p_+.
\label{sigmasol} 
\ee
Remark that the effective mass becomes $|\lambda\mu|$.


On the solution \eqref{sigmasol} the effective action \eqref{seff} can be evaluated by using the dimensional regularization as 
\be
S_{eff}/V = -N{\lambda^2(1-|\lambda|) \over 6 \pi}|\mu|^3.
\label{seffsol}
\ee

We can also evaluate momentum distribution of U($1$) flavor charge density 
\be
\aver{n_{p_s}}:=
\int \frac{d^3q}{(2 \pi)^3} \int \frac{dp_3}{(2 \pi)} 
 \left(  \aver{ \bar \psi(p) \gamma^3 \psi(q)} - \aver{ \bar \psi(p) \gamma^3 \psi(q)} |_{\mu=0} \right), 
\label{distributiondef}
\ee
where the normalization is such that it vanishes when the chemical potential does. 
By using \eqref{propagator} and \eqref{sigmasol} this becomes
\be
\aver{n_{p_s}}= 
N \left(  1 - \theta( \sqrt{ p_s^2 + (\lambda \mu)^2}  - \mu) - \theta( -\sqrt{ p_s^2 + (\lambda \mu) ^2}  - \mu ) \right).
\label{distribution}
\ee
This momentum distribution indicates this system is described by an ideal Fermi gas whose Fermi surface is $p_s = |\mu| \sqrt{1-\lambda^2}$.

To summarize, the Chern-Simons-fermion vector model in presence of chemical potential $\mu$ is effectively described by a free fermion system with the mass squared $(\lambda\mu)^2$ under the 't Hooft limit.%
\footnote{ 
In this specific model we analytically see an important assumption in statistical physics that a weakly interacting many body system reduces to a free ideal one with appropriate parametrization in the thermodynamic limit \cite{Anderson:2000}. 
}


%
%

\section{Finite temperature analysis}
\label{finitemp}

In this section we study the Chern-Simons-fermion system in nonzero temperature  $T$ by placing the system on $\bR^2 \times \bS^1$ with circumfirence $\beta=1/T$. We impose anti-periodic boundary condition for $\bS^1$ direction on the fermionic field in order to study the (grand) canonical ensemble thereof.  
Since Fourier mode in the thermal direction is quantized by $p_3 = {2\pi(n+\half) \over \beta}$, we can obtain the result in nonzero temperature from the previous calculation by doing replacement such that 
\be
\int {dp_3 \over 2\pi} F(p_3) \to \frac{1} {\beta}\sum_{p_3:F} F(p_3):=\frac{1} {\beta}\sum_{n\in\bZ} F( {2\pi(n+\half) \over \beta}),
\label{replacement}
\ee
where $F(p_3)$ is an arbitrary function of the momentum in the thermal direction. 

We first solve the Schwinger-Dyson equation, which is now given by 
\beal{
\Sigma(p) =&\frac{1} {\beta}\sum_{q_3:F}  \int {d^2 q \over (2 \pi)^2} \biggl(\gamma^+ {1\over i \gamma^\mu \wt q_\mu   + \Sigma (q) }\gamma^3 - \gamma^3 {1\over i \gamma^\mu \wt q_\mu   + \Sigma (q) }\gamma^+ \biggl)  {2\pi \lambda \over i(p-q)_- }. 
\label{sdT}
}
Repeating the same calculation we obtain 
\bea
&&\Sigma_-=\Sigma_3=0, \\
&&\Sigma_I(p) = {4\pi N \over i k} \frac{1} {\beta}\sum_{p_3:F}  \int \frac{d^2q}{(2 \pi)^2}  \biggl( { i q_-  \over   (\wt q -i \Sigma(q))^2 + \Sigma_1(q)^2}  \biggl) {1 \over (p-q)_- } \\
&& \Sigma_+(p)=  {4\pi N \over i k}\frac{1} {\beta}\sum_{p_3:F}  \int \frac{d^2q}{(2 \pi)^2} \biggl( {  \Sigma_1(q) \over   (\wt q -i \Sigma(q))^2 + \Sigma_1(q)^2}  \biggl)  {1 \over (p-q)_- }
\eea
From the dimensional ground the ansatz is such that 
\bea
\Sigma_I(p) = f(\hat p,\hat\mu) p_s, \quad \Sigma_+ (p) =i g(\hat p,\hat\mu) p_+
\label{sigmaT}
\eea
where $f$ and $g$ are unknown functions of $\hat p = {p_s \over T}$ and $\hat \mu = {\mu \over T}$. Hereafter we will suppress $\hat\mu$ in the argument. Then the above equations become
\bea
f(\hat p) p_s &=& 4\pi \lambda\frac{1} {\beta}\sum_{p_3:F} \int \frac{d^2q}{(2 \pi)^2} \biggl( {  q_-  \over   (1+g(\hat q)) q_s^2 +\wt q _3^2 + (f(\hat q) q_s ) ^2}  \biggl) {1 \over (p-q)_- },
\label{fT}\nn
g(\hat p) p_+ &=& - 4\pi \lambda\frac{1} {\beta}\sum_{p_3:F} \int \frac{d^2q}{(2 \pi)^2}  \biggl( {   f(\hat q) q_s \over  (1+g(\hat q)) q_s^2  + q_3^2 + (f(\hat q) q_s)^2}  \biggl)  {1 \over (p-q)_- }. 
\label{gT}
\eea
The same argument in the previous section leads to
\be
f(\hat p)^2 + g(\hat p) = {c \over \hat p^2}, 
\label{fgcT}
\ee
where $c$ is a dimensionless constant depending on $\lambda$ and $\hat
\mu$. 
Substituting \eqref{fgcT} into \eqref{fT} we obtain 
\beal{
f(\hat p) p_s =& 4\pi \lambda \frac{1} {\beta}\sum_{p_3:F} \int {d^2 q \over (2 \pi)^2} \biggl( {  q_-  \over q_s^2 + \wt q _3^2 + c T^2}  \biggl) {1 \over (p-q)_- }\nn
 =& 4\pi \lambda\int_{p_s}^\infty {q_s d q_s \over 2 \pi} \frac{1} {\beta}\sum_{p_3:F} {  -1 \over q_s^2 + \wt q _3^2 + c T ^2}. 
}
Regularizing UV divergence by dimensional regularization we find%
\footnote{
A useful formula to derive this is 
\bea
{1\over\beta}\sum_{n\in\bZ} {1 \over ({2\pi(n+\half) \over\beta} + i\mu)^2+h ^2 }={ \tanh{\beta(h+\mu)\over2} +  \tanh {\beta(h-\mu)\over 2} \over 4 h}. 
\eea
}
\beal{
f(\hat p) 
= { \lambda \over \hat p} \log\left(2(  \cosh \sqrt{|\hat p^2+c|} + \cosh \hat \mu) \right).\label{f2T}
}

$c$ can be determined in the same way as in the previous section. 
Substituting \eqref{f2T} into \eqref{fgcT} and expanding $g$ around $\hat p\sim0$ we find  
\be
g(\hat p) \sim {c \over \hat p^2} -  { \lambda^2 \( \log\left(2(  \cosh \sqrt{|c|} + \cosh \hat \mu) \right) \)^2 \over \hat p^2}.  
\label{gp0}
\ee
We can show $g(\hat p)\sim0$ around $p'\sim0$ from \eqref{gT} and \eqref{f2T}. 
Therefore \eqref{gp0} implies $c\geq0$ and 
\bea
\sqrt{c} &=& | \lambda | \log \left(2 ( \cosh \sqrt{c} + \cosh \hat \mu )  \right).
\label{c}
\eea
A solution exists when $|\lambda|<1$ and does not when $|\lambda|>1$. 
This fact can be easily seen by rewriting \eqref{c} as 
\be
e^{\sqrt c \over |\lambda|} = e^{\sqrt c} +e^{-\sqrt c} +e^{\hat\mu} + e^{-\hat\mu},  
\label{c2}
\ee
which suggests $e^{\sqrt c \over |\lambda| }> e^{\sqrt c}$ and thus $|\lambda|<1$.

Let us summarize the solution of the Schwinger-Dyson equation
\beal{
f(\hat p) &=  { \lambda \over \hat p}  \log  \left( 2(\cosh \sqrt{\hat p^2 +c} +  \cosh\hat\mu)\right),\quad
g(\hat p) = { c \over \hat p^2} - f(\hat p)^2,
\label{solT}
}
where $\sqrt c$ is determined by \eqref{c} or \eqref{c2}. 
Note that $\sqrt c$ physically means the thermal mass of the fermion divided by $T$.

As in the previous section 
we can compute the momentum distribution of U(1) charge density in finite temperature 
\beal{
\aver{n_{p_s}}&=
 {1\over \beta} \sum_{p_3, q_{3}:F} \int \frac{d^2q}{(2 \pi)^2} 
\left(  \aver{ \bar \psi(p) \gamma^3 \psi(q)} - \aver{ \bar \psi(p) \gamma^3 \psi(q)} |_{\mu=0} \right)
\label{distributionT}
}
where we did replacement \eqref{replacement} in \eqref{distributiondef}.  
Substituting the above solution and simplifying it we find
\beal{
\aver{n_{p_s}}&= 
{N} { - \tanh\frac{ \sqrt{ \hat p^2 + c}-\hat\mu}{2}+ \tanh\frac{\sqrt{ \hat p^2 + c}+\hat\mu}{2} \over 2}. 
\label{distributionT}
}
This is the momentum distribution of the free fermion gas in nonzero temperature with thermal mass  $\sqrt cT$. The fermi surface clearly seen in \eqref{distribution} is now smeared by finite temperature effect.

%
%

\begin{figure}
  \begin{center}
  \subfigure[]{\includegraphics[scale=.5]{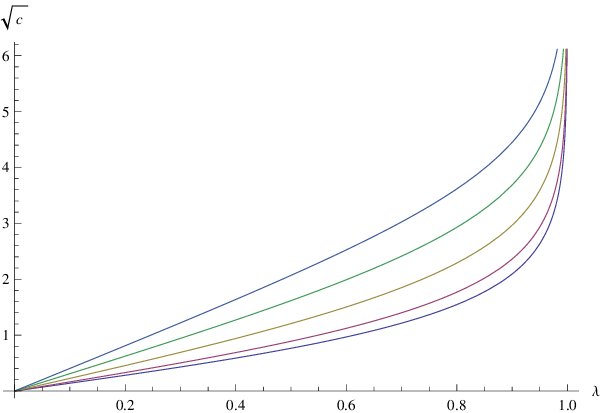}
\label{sqrtc}
  }
  \qquad\qquad
  \subfigure[]{\includegraphics[scale=.5]{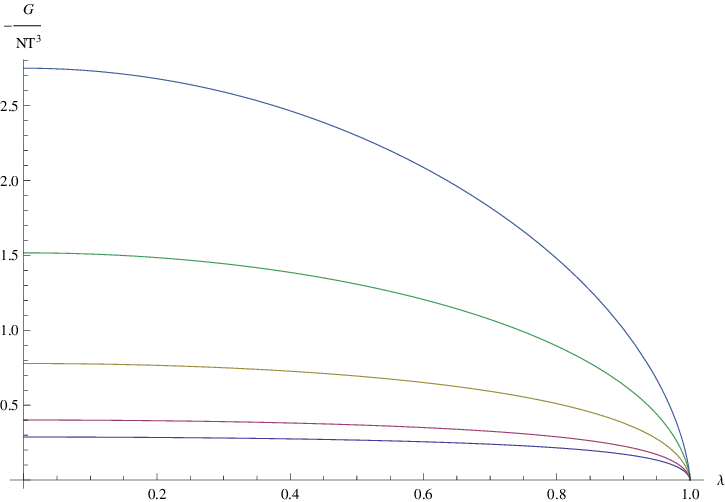}
\label{grandpot}
  }
  \end{center}
  \vspace{-0.5cm}
  \caption{The graphs of $\sqrt c$ and $-{G(T,\mu)\over NT^3}$ are shown as a function of $\lambda$ varying $\mu/T = 0,1,2,3,4$ from the bottom in Fig.\ref{sqrtc}, Fig.\ref{grandpot}, respectively. They are  taking finite values in the region $0\leq\lambda <1$. }
  \label{}
\end{figure}

Let us compute the grand canonical potential density in this system, which is defined by 
\bea
G(T,\mu)= \frac{S_{eff}-S_{eff}|_{T,\mu=0}}{V_2 \beta}.
\eea
where $V_2$ is the volume of two plane given by $V_2 = V/\beta$ and $S_{eff}$ is the effective action on $\bR^2 \times \bS^1$ with chemical potential 
\begin{equation}\label{sfta}
S_{eff} 
= NV_2 \sum_{q_3:F} \int \frac{d^2 q}{(2 \pi)^2} {\rm tr}
\left[-\log\left[ i \gamma^\mu \wt q_\mu + \Sigma(q) \right] 
+\frac{1}{2} \Sigma(q)
 \left( \frac{1}{i \gamma^\mu \wt q_\mu + \Sigma(q)} \right)  \right]
\end{equation}
and we normalize the grand canonical potential as zero when $T=\mu=0$. 
By employing the dimensional regularization the grand canonical potential density can be computed as%
\footnote{
We note a useful formula 
\begin{align}
&{1\over\beta} \sum_{n \in \bZ} \log\left(({2\pi(n+\half)\over \beta} + i \mu )^2 + h^2\right) = \log (1 + e^{ - \beta (h+\mu)} ) +  \log (1 + e^{ - \beta (h-\mu)} ) +h. 
\end{align}
}  
\begin{align}
G(T,\mu) &= 
- \frac{N T^3}{6 \pi} \biggl( 
\frac{\sqrt c^3 }{|\lambda|} 
-  \sqrt c^3 + 3\int_{\sqrt c}^\infty dy ~ y \ln \left ( 1+e^{-y-\hat\mu} \right) 
+ 3 \int_{\sqrt c}^\infty dy ~  y \ln \left ( 1+e^{-y+\hat\mu} \right) \biggr). 
\label{gcpd}
\end{align}
We plot $\sqrt c$ and $-{G(T,\mu)\over NT^3}$ as a function of $\lambda$ varying $\hat\mu$ in Fig.\ref{sqrtc}, Fig.\ref{grandpot}, respectively. We have assumed $\lambda\geq0$ here.
Those figures indicate that as $|\lambda|$ goes larger the thermal mass becomes bigger and the gratitude of the grand canonical potential smaller, and finally when $|\lambda|$ reaches one the thermal mass blows up and the $\cO(N)$ grand canonical potential vanishes. 
Note that this grand canonical potential \eqref{gcpd} reduces to the free energy density obtained in \cite{Giombi:2011kc} when $\mu\to0$ and to the effective action \eqref{seffsol} when $T\to0$ since $\sqrt c \to |\lambda\mu|/T$ from \eqref{c}. 

The behaviors of $\sqrt c$ and $G(T,\mu)$ around $\hat\mu\sim\infty$, which means low temperature or small chemical potential, with $\lambda$ fixed is the following.
\beal{
\sqrt c &\sim \lambda\hat\mu + \lambda e^{(\lambda-1)\hat\mu} + \cdots,\\
G(T,\mu) &\sim  
- \frac{N T^3}{6 \pi} \biggl( 
 \half(1 - \lambda^2) \hat\mu^3 + {\pi^2 \over 2} \hat\mu +\cdots\biggr).
}
Their behaviors around $\lambda \sim 1$ with ${\hat\mu}$ fixed are
\beal{
\sqrt c &\sim \log \left( { 2 \cosh {\hat\mu} \over 1 -\lambda} \right) 
- \log \log \left( { 2 \cosh {\hat\mu} \over 1 -\lambda} \right) +\cdots,  \\
G(T,\mu) &\sim
- \frac{N T^3 ( 1-\lambda) }{6 \pi} \biggl( \sqrt c^3 + {3 \over  \cosh {\hat\mu}} \sqrt c^2 + \cdots \biggr).
}


From the grand canonical potential one can obtain its thermodynamical descendants such as the pressure $p$, the mean U($1$) flavor charge density $Q$, the average internal energy $U$, the entropy density $S$ and heat capacitance $C$ with fixed two dimensional volume.
\bea
p=-G,\;
Q= - (\partial_\mu G)_{{T}},\;
U= \partial_{\beta}( \beta G)_{{\hat\mu}},\;
S = - \left( {\partial  G \over \partial T}\right)_{{\mu}}, \;
C= -T \left( {\partial ^2 G \over \partial T^2}\right)_{{\mu}},
\eea
where subscripts are such that partial derivatives are taken with them fixed. 
From a straight forward calculation we obtain 
\bea
Q =  {N T^2 \over 6 \pi} g', \quad
U=-2 G,\quad
S=  {N T^2 \over 2 \pi} (g  - {{\hat\mu} \over 3} g' ), \quad
C =  {N T^2 \over \pi} ( g - {2 \over 3} {\hat\mu} g '+{ {\hat\mu}^2 \over 6} g''),
\eea
where we set 
$g=- { 6 \pi \over N T^3} G(T,\mu)=g({\hat\mu}) $ and $g'={d g \over d \hat\mu}, g''={d^2 g \over d \hat\mu^2}$.%
\footnote{ The explicit forms of $g'({\hat\mu}), g''({\hat\mu})$ are calculated as
\beal{ 
g'({\hat\mu}) &= 3 \biggl[ \sqrt c \log
\left( { \cosh\frac{\sqrt c - {\hat\mu}}{2} \over \cosh\frac{\sqrt c + {\hat\mu}}{2} } \right) 
+ \int _{\sqrt c -{\hat\mu}}^{\sqrt c + {\hat\mu}} dy  \log (2 \cosh \frac{y}{2})\biggr],\\
 g''({\hat\mu}) &= 3 \biggl[
{\sqrt c \over 2}  (\partial_{\hat\mu} \sqrt c -1) \left( \tanh\frac{\sqrt c - {\hat\mu}}{2} -  \tanh\frac{\sqrt c + {\hat\mu}}{2}  \right) 
+ {\sqrt c \over \lambda} \biggr].
} 
}
In particular 
we can show $U$ and $Q$ are related in large $\hat\mu$ region by
$
U \sim  \sqrt{ 16 \pi \over 9 N ( 1 -\lambda^2)} {Q} ^\frac{3}{2}.
$

\begin{figure}
  \begin{center}
  \subfigure[]{\includegraphics[scale=.3]{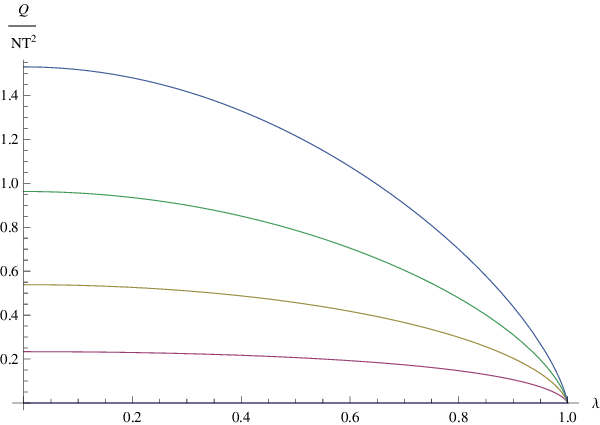}
\label{charge}
  }
  \qquad\qquad
  \subfigure[]{\includegraphics[scale=.3]{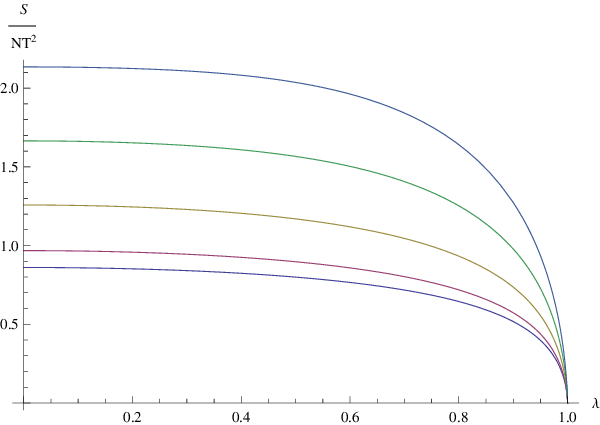}
\label{entropy}
  }
  \qquad\qquad
  \subfigure[]{\includegraphics[scale=.3]{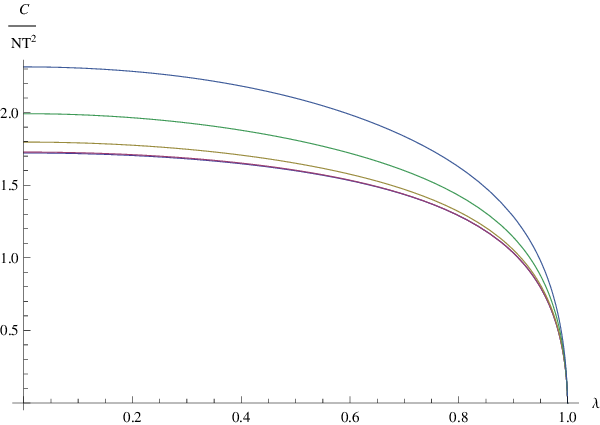}
\label{heatcap}
  }
  \end{center}
  \vspace{-0.5cm}
  \caption{The U(1) flavor charge density, entropy density and heat capacitance normalized by $NT^2$ are plotted as a function of $\lambda$ varying $\mu/T = 0,1,2,3,4$ from the bottom in Fig.\ref{charge}, Fig.\ref{entropy}, Fig.\ref{heatcap}, respectively. }
  \label{}
\end{figure}

\section{Comment on instability in $|\lambda|>1$} 
\label{instability}

In the previous section we solved the Chern-Simons-fermion vector theory with chemical potential in finite temperature in the 't Hooft limit. The solution exists when $|\lambda|<1$ and ceases to exist at $|\lambda|=1$. It is natural to ask what happens when $|\lambda|>1$. 
It is likely that at $|\lambda|=1$ the system undergoes the phase transition from conformally symmetric (deconfining) phase in $|\lambda|<1$ to broken (confining) one in $|\lambda|>1$ under the assumption that the theory exists beyond $|\lambda|=1$.%
\footnote{It is also possible that the theory does not exist beyond $|\lambda|=1$ since the regularization by the Yang-Mills term requires $|\lambda|$ to be less than one.}
Let us briefly discuss the possibility of this phase transition by employing the results obtained above. 

For this purpose we consider the same system in the U($N$) (not SU($N$)) gauge group and
take the diagonal U($1$) holonomy in the thermal direction into account, 
which has the leading contribution in $1/N$ expansion. 
The self-energy and free energy density including the holonomy contribution can be respectively obtained directly from the self-energy \eqref{sigmaT} with \eqref{solT}, \eqref{c2} and the grand canonical potential density \eqref{gcpd} by performing an analytic continuation for chemical potential 
$
\mu \to i\alpha, 
$
where $\alpha$ represents the diagonal U($1$) holonomy. 
After this manipulation the solution of the self energy becomes 
\beal{
f(\hat p) &=  { \lambda \over \hat p}  \log  \left( 2(\cosh \sqrt{\hat p^2 +c} +  \cos\hat\alpha)\right),\quad
g(\hat p) = { c \over \hat p^2} - f(\hat p)^2,\\
e^{\sqrt c \over |\lambda|} &= e^{\sqrt c} +e^{-\sqrt c} +2\cos\hat\alpha,  
\label{solch}
}
where $\hat\alpha = {\alpha \over T}$. There exists a solution in $0\leq\hat\alpha\leq\pi$ when $|\lambda|<1$. 
In the region $|\lambda|>1$ there also exists a solution in $\hat\alpha_c  \leq\hat\alpha\leq\pi$, where $\alpha_c$ is a function of $\lambda$ taking values between $\pi/2$ and $2\pi/3$.%
\footnote{
Precisely speaking, $\alpha_c$ is given by $\alpha_c(\lambda) = \arccos\text{Max}_{\sqrt c}\(\half \exp({\sqrt c \over |\lambda|}) -\cosh\sqrt c\)$, where $\text{Max}_{x} A(x)$ is the maximum value of $A(x)$ with respect to $x$.  
}
The free energy density is 
\begin{align}
F(T,\alpha) &= 
- \frac{N T^3}{6 \pi} \biggl( 
\frac{\sqrt c^3 }{|\lambda|} 
-  \sqrt c^3 + 3\int_{\sqrt c}^\infty dy ~ y \ln \left ( 1+2\cos\hat\alpha e^{-y}+e^{-2y} \right) \biggr). 
\label{feh}
\end{align}
Since the holonomy is dynamical, 
the large $N$ leading free energy is determined by varying the holonomy satisfying \eqref{solch} so that the free energy is minimized.
It turns out by simple numerics that
the value of the holonomy which minimizes the free energy is trivial when $|\lambda|<1$,%
\footnote{
This result was assumed in \cite{Giombi:2011kc} to determine the free energy.
} 
but becomes non-trivial, $\alpha=\alpha_c$, when $|\lambda|>1$.

This result suggests that the first order phase transition may happen at $|\lambda|=1$. 
However it also turns out that 
the minimized free energy \eqref{feh} at $\alpha=\alpha_c$ in $|\lambda|>1$ is positive, 
which means the pressure is negative and thus the system is no longer stable. 
This is presumably a signal of break-down of our assumption for the system to have conformal invariance. If the theories exist they cannot be in a conformally symmetric phase beyond $|\lambda|=1$. Operators neglected in this analysis, which break conformal symmetry, would play an important role to investigate the phase in $|\lambda|>1$. 
We leave detailed study in the future.

\section{Summary and discussion} 
\label{discussion}

We have studied SU($N$) Chern-Simons theory coupled to a fermion in the fundamental representation 
with chemical potential for U($1$) flavor symmetry and temperature in the 't Hooft limit. 
We have calculated the self energy, the momentum distribution and several thermodynamic quantities  exactly as a function of $\lambda$, which suggests that the solution exists up to $|\lambda|=1$ for a fixed value of chemical potential. We have discussed instability of the system beyond $|\lambda|=1$ by considering U($N$) gauge group and using the diagonal U($1$) holonomy.

Even though the technique to solve Chern-Simons vector model seems to work well, there remains a puzzle against the result of calculation of three point functions \cite{Maldacena:2011jn,Maldacena:2012sf,Aharony:2012nh}. 
The result of three point functions suggests that there exists a continuous parameter (just like the 't Hooft coupling) which interpolates between free/Gross-Neveu fermion vector models and critical/free scalar vector models respectively. However it seems this ``bosonization'' in three dimensions is not seen from the free energies computed by using the technique in \cite{Giombi:2011kc,Jain:2012qi}.  
We do not know the answer of this puzzle.

Related to this issue it is quite interesting to study the sub-leading correction in $1/N$ expansion since
it might be possible that $1/N$ expansion breaks down in such a limit as pointed out in \cite{Lee:2009,Aharony:2012nh}. In that case the free energy obtained in \cite{Giombi:2011kc,Jain:2012qi} would be modified so that non-planer diagrams are taken into account. At least we expect 
the next-leading contribution will be obtained by considering the contribution of the ladder diagrams, 
which will be technically much harder but considerably important to compute.

We hope to come back to these issues in the future.

\section*{Acknowledgments} 

The author thanks Sachin Jain, Shiroman Prakash, Nilanjan Sirkar, Sandip P. Trivedi, Spenta R. Wadia and especially  Shiraz Minwalla for collaboration and discussion on various stages of this work. 
He is also grateful to Sandip P. Trivedi and Spenta R. Wadia for reading the manuscript and giving helpful comments.

\bibliographystyle{utphys}
\bibliography{ref}

\end{document}